\documentclass[prd,preprint,showpacs,amsmath,amssymb,floatfix]{revtex4}
\usepackage{graphicx,color,dcolumn}
\begin{document}
\title{$B_{s1}(5830)$ and $B_{s2}^*(5840)$}
\author{Zhi-Gang Luo$^1$
}
\email{zgluo@pku.edu.cn}
\author{Xiao-Lin Chen$^1$
}
\author{Xiang
Liu$^{2,1}$
\footnote{Corresponding
author}}\email{liuxiang@teor.fis.uc.pt}
\affiliation{$^1$School of Physics, Peking University, Beijing 100871, China\\
             $^2$Centro de F\'{i}sica Computacional, Departamento de F\'{i}sica, Universidade de Coimbra,
             P-3004-516, Coimbra, Portugal}
\date{\today}
\begin{abstract}
In this paper we investigate the strong decays of the two newly
observed bottom-strange mesons $B_{s1}(5830)$ and $B_{s2}^*(5840)$
in the framework of the quark pair creation model. The two-body
strong decay widths of $B_{s1}(5830)^0\to B^{*+}K^-$ and
$B_{s2}^*(5840)^0\to B^+K^-,\,B^{*+}K^-$ are calculated by
considering $B_{s1}(5830)$ to be a mixture between $|^1P_1\rangle$
and $|^3P_1\rangle$ states, and $B_{s2}^*(5840)$ to be a
$|^3P_2\rangle$ state. The double pion decay of $B_{s1}(5830)$ and
$B_{s2}^*(5840)$ is supposed to occur via the intermediate state
$\sigma$ and $f_0(980)$. Although the double pion decay widths of
$B_{s1}(5830)$ and $B_{s2}^*(5840)$ are smaller than the two-body
strong decay widths of $B_{s1}(5830)$ and $B_{s2}^*(5840)$, one
suggests future experiments to search the double pion decays of
$B_{s1}(5830)$ and $B_{s2}^*(5840)$ due to their sizable decay
widths.
\end{abstract}

 \maketitle

\section{Introduction}\label{sec1}

Heavy flavor physics is an interesting research field. In the past
three years, a series of the new observations of the heavy flavor
hadrons, such as $D_{sJ}(2317)$, $D_{sJ}(2460)$
\cite{Aubert:2003fg,Krokovny:2003zq,Besson:2003cp,others},
$D_{sJ}(2860)$ \cite{2860-babar}, $D_{sJ}(2715)$
\cite{belle-2715,Belle-2715}, $\Lambda_{c}(2880,2940)^+$,
$\Xi_c(2980,3077)^{+,0}$, $\Omega_{c}(2768)^0$
\cite{babar-2880,belle-2880,babar-2980-3077,belle-2980-3077,new-Xi3055,babar-omega,cleo-2880},
$\Sigma_{b}^{\pm}$, $\Sigma_{b}^{*\pm}$ \cite{CDF,CDF-1}, and
$\Xi_b$ \cite{D0-Xi,CDF-Xi,CDF-Xi-1}, have made the study of heavy
flavor physics active and attractive.

\begin{center}
 \begin{figure}[htb]
 \scalebox{0.8}{\includegraphics{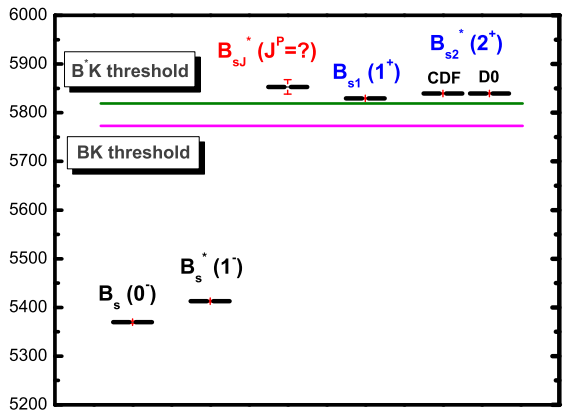}}\caption{The mass
 spectrum of bottom-strange mesons. The data is taken from particle date group (PDG) \cite{PDG} and the CDF
 and D0 experiments \cite{CDF-Bs,D0-Bs}.} \label{spectrum}
 \end{figure}
\end{center}

Up to now, there only exist two established bottom-strange mesons
in Particle Data Group (PDG) \cite{PDG}. However, recent
observations of the two orbitally excited $B_s$ mesons announced
by CDF \cite{CDF-Bs,D0-Bs} and D0 experiments make the
bottom-strange mass spectrum become abundant. The CDF
collaboration reported $m_{B_{s1}}=5829.4\pm0.7$ MeV and
$m_{B_{s2}^*}=5839.6\pm 0.7$ MeV \cite{CDF-Bs}. The D0
collaboration confirmed $B_{s2}^{*}(5840)$ state with
$m_{B_{s2}^*}=5839.6\pm 1.1(\mathrm{stat}.)\pm
0.7(\mathrm{syst}.)$ MeV \cite{D0-Bs}, and indicated that
$B_{s1}(5830)$ was not observed with the available data set
\cite{D0-Bs}. In Fig. \ref{spectrum}, one lists all bottom-strange
mesons observed by the experiments.

For heavy-light meson system, we can group it into several
doublets in terms of the heavy quark effective theory (HQET), i.e.
$j_{\ell}^P=\frac{1}{2}^-$ $H$ doublet $(0^-, 1^-)$ with orbital
angular momentum $L=0$, $j_{\ell}^P=\frac{1}{2}^+$ $S$ doublet
$(0^+,1^+)$ and $j_{\ell}^P=\frac{3}{2}^+$ $T$ doublet $(1^+,2^+)$
with $L=1$. The D0 and CDF experiments indicated that
$B_{s1}(5830)$ and $B_{s2}^*(5840)$ correspond to the states
respectively with $J^P=1^+$ and $J^{P}=2^+$ in $T$ doublet
\cite{CDF-Bs,D0-Bs}.

Before finding $B_{s1}(5830)$ and $B_{s2}^*(5840)$, many
theoretical groups were involved in the study of the properties of
heavy-light mesons. In Ref. \cite{godfrey}, the authors studied
the masses of P-wave states by the relativistic quark model, then
calculated their decay widths using both the pseudoscalar emission
model and the flux-tube-breaking model. Eichten, Hill and Quigg
estimated the masses and the decay widths of orbitally excited
heavy-light mesons by using the heavy quark symmetry, which is
supplemented by the insights from the potential model \cite{EHQ}.
Ebert, Galkin and Faustov calculated the mass spectrum of the
orbitally excited heavy-light mesons according to the relativistic
quark model \cite{EGF}. Then Di Pierro and Eichten carried out a
detailed study of the orbital and radial excited heavy mesons
\cite{DE}. By the effective Lagrangian constructed in the chiral
symmetry and the heavy quark limit, Falk and Mehen examined the
decays of the excited heavy mesons including the leading power
corrections to the heavy quark limit \cite{FM}. In the approach of
Lattic QCD, the authors of Ref. \cite{Green} obtained the mass
spectrum of the excited heavy-light meson. In Ref.
\cite{colangelo}, Colangelo, Fazio and Ferrandes studied the
structures and the decays of the orbitally excited states.
Matsuki, Morii and Sudoh obtained the mass spectrum of the
heavy-light systems by the semi-relativistic quark model
\cite{Matsuki}. According to the chiral quark model, Zhong and
Zhao performed the calculations of the strong decays of the
heavy-light mesons \cite{zhong}. All of the above mentioned work
refers to the bottom-strange mesons.

The observations of the two bottom-strange states have inspired
our interest in $B_{s1}(5830)$ and $B_{s2}^*(5840)$, especially in
their decay properties. In Ref. \cite{semi-liu}, one performed the
calculations of the semileptonic decays of $B_{s1}(5830)$ and
$B_{s2}^*(5840)$. At present, the CDF and D0 experiments only
carried out the measurements of the masses of $B_{s1}(5830)$ and
$B_{s2}^*(5840)$. However, the total widths of $B_{s1}(5830)$ and
$B_{s2}^*(5840)$ are still missing. Thus the study on their strong
decay becomes an interesting and important topic, which will be
helpful not only for obtaining the information of the total widths
of $B_{s1}(5830)$ and $B_{s2}^*(5840)$, but also for testing the
model applied to the calculation of the strong decay of
$B_{s1}(5830)$ and $B_{s2}^*(5840)$. In this work, we focus on the
calculation of the strong decay rates of $B_{s1}(5830)$ and
$B_{s2}^*(5840)$ using the $^3P_0$ model.

This work is organized as follows. After the introduction, we
briefly review the $^3P_0$ model. In Sec. \ref{sec3} and Sec.
\ref{sec4}, we present the formulation and the numerical result of
the two-body and double pion decays of $B_{s1}(5830)$ and
$B_{s2}^*(5840)$, respectively. The last section is a short
summary.

\section{A review of the $^3P_0$ model}\label{sec2}

In this work we use the $^3P_0$ model
\cite{Micu,yaouanc,yaouanc-1,yaouanc-book,Beveren,BSG,sb}, also
known as the Quark Pair Creation (QPC) model, to calculate the
strong decays of $B_{s1}(5830)$ and $B_{s2}^*(5840)$. This model
is applicable to Okubo-Zweig-Iizuka (OZI) allowed strong decays of
a hadron into two other hadrons, which are expected to be the
dominant decay modes of a meson if they are allowed. The $^3P_0$
model has been widely used since it is successful when applied
extensively to the calculation of the strong decay of hadron
\cite{qpc-1,qpc-2,qpc-90,ackleh,Zou,liu,Close:2005se,lujie,xiangliu-2860,xiangliu-heavy,Li:2008mz}.

\begin{center}
\begin{figure}[htb]
\scalebox{0.7}{\includegraphics{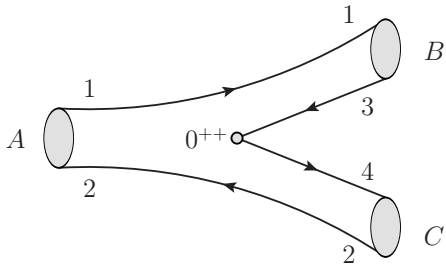}} \caption{The $^3P_0$
decay mechanism for meson decay $A\to B+C$.} \label{decay}
\end{figure}
\end{center}
In the QPC model, the heavy meson decay occurs via a
quark-antiquark pair production from the vacuum, which is depicted
in Fig. \ref{decay}. The created quark pair is of the quantum
number of the vacuum, $0^{++}$ \cite{Micu,yaouanc}. In the
non-relativistic limit, the transition operator is expressed as
\begin{eqnarray}
T&=& - 3 \gamma \sum_m\: \langle 1\;m;1\;-m|0\;0 \rangle\,
\int\!{\rm d}{\textbf{k}}_3\; {\rm d}{\textbf{k}}_4
\delta^3({\textbf{k}}_3+{\textbf{k}}_4) {\cal
Y}_{1m}\left(\frac{{\textbf{k}}_3-{\textbf{k}_4}}{2}\right)\;
\nonumber\\&&\times\chi^{3 4}_{1, -\!m}\; \varphi^{3 4}_0\;\,
\omega^{3 4}_0\; d^\dagger_{3i}({\textbf{k}}_3)\;
b^\dagger_{4j}({\textbf{k}}_4)\,, \label{tmatrix}
\end{eqnarray}
where $i$ and $j$ are the SU(3)-color indices of the created quark
and anti-quark. $\varphi^{34}_{0}=(u\bar u +d\bar d +s \bar
s)/\sqrt 3$ and $\omega_{0}^{34}=\delta_{ij}$ are for flavor and
color singlets, respectively. $\chi_{{1,-m}}^{34}$ is a triplet
state of spin. $\mathcal{Y}_{\ell m}(\mathbf{k})\equiv
|\mathbf{k}|^{\ell}Y_{\ell m}(\theta_{k},\phi_{k})$ is the
$\ell$th solid harmonic polynomial. $\gamma$ is a dimensionless
constant which denotes the strength of quark pair creation from
vacuum and can be extracted by fitting data. We adopt the mock
state to describe the meson with the spatial wave function
$\Psi_{n_A L_A M_{L_A}}\left(\mathbf{k}_1,\mathbf{k}_2\right)$ in
the momentum representation \cite{mockmeson} {\small
\begin{eqnarray}\label{mockmeson}
&&\left|A(n_A \mbox{}^{2S_A+1}L_A \,\mbox{}_{J_A M_{J_A}})
({\textbf{K}}_A) \right\rangle \nonumber\\&&= \sqrt{2
E_A}\sum_{M_{L_A},M_{S_A}} \left\langle L_A M_{L_A} S_A M_{S_A} |
J_A M_{J_A} \right\rangle \nonumber\\&&\quad\times\int \rm d
\mathbf{k}_1\rm
d\mathbf{k}_2\delta^3\left(\textbf{K}_A-\mathbf{k}_1-{\mathbf{k}}_2\right)\Psi_{n_A
L_A M_{L_A}}\left(\mathbf{k}_1,\mathbf{k}_2\right) \nonumber\\
&&\quad\times\chi^{1 2}_{S_A M_{S_A}}\varphi^{1 2}_A\omega^{1 2}_A
\left|\;q_1\left(\mathbf{k}_1\right)
\bar{q}_2\left(\mathbf{k}_2\right)\right\rangle,
\end{eqnarray}}
which satisfies the normalization conditions
\begin{eqnarray}\label{nor-cond}
\langle A(\textbf{K}_A)|A(\textbf{K}'_A) \rangle &=& 2E_A\,
\,\delta^3(\textbf{K}_A-\textbf{K}'_A),\;\\
\langle
q_i(\mathbf{k}_i)|q_j(\mathbf{k}_j)\rangle&=&\delta_{ij}\delta^3(\mathbf{k}_i-\mathbf{k}_j),\;\\
\langle
\bar{q}_i(\mathbf{k}_i)|\bar{q}_j(\mathbf{k}_j)\rangle&=&\delta_{ij}\delta^3(\mathbf{k}_i-\mathbf{k}_j),\;
\end{eqnarray}
\begin{eqnarray}
&&\int \rm d\mathbf{k}_1 \rm d\mathbf{k}_2
\delta^3(\textbf{K}_A-\mathbf{k}_1-\mathbf{k}_2)
\Psi_A(\mathbf{k}_1,\mathbf{k}_2)\Psi_{A'}(\mathbf{k}_1,\mathbf{k}_2)=\delta_{A'A}.
\end{eqnarray}
The subscripts 1 and 2 in (\ref{mockmeson}) refer to the quark and
the anti-quark within the meson $A$, respectively.
${\textbf{K}}_A$ is the momentum of the meson $A$.
$\mathbf{S}_A=\mathbf{s}_{q_1}+\mathbf{s}_{q_2}$ is the total
spin. $\mathbf{J}_A=\mathbf{L}_A+\mathbf{S}_A$ denotes the total
angular momentum.

For $A\to B+C$ process, the S-matrix is depicted as
\begin{eqnarray}
\langle
BC|S|A\rangle=I-i2\pi\delta(E_f-E_i)\langle{}BC|T|A\rangle.
\end{eqnarray}
In the center of the mass frame of the meson $A$, $\textbf{K}_A=0$
and $\textbf{K}_B=-\textbf{K}_C=\textbf{K}$. Then, we have
\begin{eqnarray}\label{T-matrix}
&&\langle BC|T|A\rangle=\sqrt{8 E_A E_B
E_C}\;\;\gamma\!\!\!\!\!\!\!\!\!\!\!
\sum_{\renewcommand{\arraystretch}{.5}\begin{array}[t]{l}
\scriptstyle M_{L_A},M_{S_A},\\
\scriptstyle M_{L_B},M_{S_B},\\
\scriptstyle M_{L_C},M_{S_C},m
\end{array}}\renewcommand{\arraystretch}{1}\!\!\!\!\!\!\!\!
\langle 1\;m;1\;-m|\;0\;0 \rangle\nonumber\\&&\quad\times \langle
L_A M_{L_A} S_A M_{S_A} | J_A M_{J_A} \rangle\langle L_B M_{L_B}
S_B M_{S_B} | J_B M_{J_B} \rangle\nonumber\\&&\quad\times\langle
L_C M_{L_C} S_C M_{S_C} | J_C M_{J_C} \rangle\langle\varphi^{1
3}_B \varphi^{2 4}_C | \varphi^{1 2}_A \varphi^{3 4}_0
 \rangle\nonumber\\&&\quad\times
 \langle \chi^{1 3}_{S_B M_{S_B}}\chi^{2 4}_{S_C
M_{S_C}}  | \chi^{1 2}_{S_A M_{S_A}} \chi^{3 4}_{1 -\!m} \rangle
I^{M_{L_A},m}_{M_{L_B},M_{L_C}}({\textbf{K}}) \;.\nonumber\\
\end{eqnarray}
The spatial integral $I^{M_{L_A},m}_{M_{L_B},M_{L_C}}(\textbf{K})$
reads as
\begin{eqnarray}
&&I^{M_{L_A},m}_{M_{L_B},M_{L_C}}(\textbf{K}) = \int\!\rm
d\mathbf{k}_1\rm d\mathbf{k}_2\rm d\mathbf{k}_3\rm
d\mathbf{k}_4\,\delta^3(\mathbf{k}_1+\mathbf{k}_2)\nonumber\\
&&\quad\times\delta^3(\mathbf{k}_3+\mathbf{k}_4)\delta^2
(\textbf{K}_B-\mathbf{k}_1-\mathbf{k}_3)\delta^3(\textbf{K}_C-\mathbf{k}_2-\mathbf{k}_4)\nonumber\\
&&\quad\times\Psi^*_{n_B L_B
M_{L_B}}(\mathbf{k}_1,\mathbf{k}_3)\Psi^*_{n_C L_C
M_{L_C}}(\mathbf{k}_2,\mathbf{k}_4)\nonumber\\&&\quad\times
\Psi_{n_A L_A M_{L_A}}(\mathbf{k}_1,\mathbf{k}_2)
\mathcal{Y}_{1m}\Big(\frac{\mathbf{k}_3-\mathbf{k}_4}{2}\Big).
\label{integral}
\end{eqnarray}
The rest of the model is just to describe the overlap of the
initial meson ($A$) and the created pair with the two final mesons
($B$ and $C$), and then finally to calculate the probability that
the rearrangement will occur. The radial portions of the meson
space wavefunction can be expressed in certain functional forms,
which encompass the simple harmonic oscillator (HO) wavefunction
\begin{eqnarray}
\Psi_{nLM}(\mathbf{k})=\mathcal{N}_{nL}\exp\left(-\frac{R^2\mathbf{k}^2}{2}\right)\mathcal{Y}_{LM}(\mathbf{k})\,\mathcal{P}(\mathbf{k}^2),
\end{eqnarray}
where $\mathcal{P}(\mathbf{k}^2)$ is the polynomial of
$\mathbf{k}^2$. $\mathbf{k}$ is the relative momentum between the
quark and the anti-quark within a meson. For example, meson $A$ is
composed of quark $1$ and anti-quark $2$, so,
$\mathbf{k}_A=(m_2\mathbf{k}_1-m_1\mathbf{k}_2)/(m_1+m_2).$
$\mathcal{N}_{nL}$ denotes the normalization coefficient. In this
work, for the decay channels of interest, what we need is only the
lowest two states without the radical excitation, i.e.
\begin{eqnarray}
\Psi_{00}(\mathbf{k})&=&\frac{1}{\pi^{3/4}}R^{3/2}\exp\left(-\frac{R^2\mathbf{k}^2}{2}\right),\\
\Psi_{1\mu}(\mathbf{k})&=&i\frac{\sqrt{2}}{\pi^{3/4}}R^{5/2}k_\mu\exp\left(-\frac{R^2\mathbf{k}^2}{2}\right),
\end{eqnarray}
where $k_\mu$ is the spherical component of the vector
$\mathbf{k}$, which is defined as
$k_{\pm1}=\mp(k_x\pm{}ik_y)/\sqrt{2}$ and $k_{0}=k_z$.

In terms of Wigner's $9j$ symbol, the spin matrix element can be
written as \cite{yaouanc-book}
\begin{eqnarray}
&&{ \langle  \chi^{1 3}_{B_C M_{B_C}} \chi^{2 4}_{S_C M_{S_C}}|
\chi^{1 2}_{S_A M_{S_A}} \chi^{3 4}_{1 -\!m} \rangle}
\nonumber\\\quad&& =(-1)^{S_{C}+1}
\Big{[}3(2S_B+1)(2S_C+1)(2S_A+1)\Big{]}^{1/2}\nonumber\\&&\quad\times\sum_{S,M_s}\langle
S_BM_{S_B}S_CM_{S_C} |SM_s \rangle \nonumber\\&&\quad\times\langle
SM_s|S_AM_{S_A};1, -m \rangle\; \left \{\begin{array}{ccc}
1\over 2 & 1\over 2 & S_B \nonumber\\
1\over 2 & 1\over 2 & S_C\\
S_A & 1 & S
\end{array}
\right \} \;.
\end{eqnarray}

With the transition amplitude obtained in (\ref{T-matrix}), the
helicity amplitude $\mathcal{M}^{M_{J_A}M_{J_B}M_{J_C}}$ can be
extracted from
\begin{eqnarray}
\langle{}BC|T|A\rangle=\delta^3(\mathbf{K}_B+\mathbf{K}_C-\mathbf{K}_A)\mathcal{M}^{M_{J_A}M_{J_B}M_{J_C}}.
\end{eqnarray}
The decay width for the process $A\to BC$ in terms of the helicity
amplitude is
\begin{eqnarray*}
\Gamma=\pi^2\frac{|\mathbf{K}|^2}{M_A^2}\frac{1}{2J_A+1}
\sum_{\renewcommand{\arraystretch}{.5}\begin{array}[t]{l}
\scriptstyle M_{J_{M_A}},M_{J_{M_B}},\\
\scriptstyle \quad M_{J_{M_C}}
\end{array}}
\Big|\mathcal{M}^{M_{J_A}M_{J_B}M_{J_C}}\Big|^2\,.
\end{eqnarray*}
For the sake of convenience, one usually works out the partial
wave amplitude first via the Jacob-Wick formula \cite{convert}
\begin{eqnarray}
&&{\mathcal{M}}^{J L}(A\rightarrow BC) = \frac{\sqrt{2 L+1}}{2 J_A
+1} \!\! \sum_{M_{J_B},M_{J_C}} \langle L 0 J M_{J_A}|J_A
M_{J_A}\rangle \nonumber\\&&\quad\quad\quad\times\langle J_B
M_{J_B} J_C M_{J_C} | J M_{J_A} \rangle \mathcal{M}^{M_{J_A}
M_{J_B} M_{J_C}}({\textbf{K}}),
\end{eqnarray}
where $\mathbf{J}=\mathbf{J}_B+\mathbf{J}_C$ and
$\mathbf{J}_{A}
=\mathbf{J}_{B}+\mathbf{J}_C+\mathbf{L}$.
Then one calculates the decay width in terms of the partial wave
amplitude
\begin{eqnarray}
\Gamma = \pi^2 \frac{{|\textbf{K}|}}{M_A^2}\sum_{JL}\Big
|\mathcal{M}^{J L}\Big|^2,\label{de}
\end{eqnarray}
where $|\textbf{K}|$, as mentioned above, is the three momentum of
the daughter mesons in the parent's center of mass frame.

\section{Two-body strong decays}\label{sec3}

The two-body strong decays of $B_{s1}(5830)^0$ and
$B_{s2}^*(5840)^0$ allowed by the phase space include
\begin{eqnarray*}
\left\{\begin{array}{l}
B_{s1}(5830)^0\to B^{*+}K^-,\;B^{*0}\bar K^0\\
B_{s2}^*(5840)^0\to B^+K^-,\;B^{0}\bar{K}^0
\\B_{s2}^*(5840)^0\to
B^{*+}K^-,\;B^{*0}\bar K^0\end{array}\right.\,.
\end{eqnarray*}
Due to the conservations of the angular momentum and the parity,
the $B\bar K$ decay mode for $B_{s1}(5830)^0$ is forbidden.

Before entering the calculation, we firstly introduce the
component of $B_{s1}(5830)^0$ with $J^{P}=1^+$. In quark model,
$B_{s1}(5830)^0$ is usually considered as the mixture of the two
basis states $|^1P_1\rangle$ and $|^3P_1\rangle$ \cite{godfrey}
\begin{eqnarray}
\left (\begin{array}{c}
 \left|1^+,j_l^P=\frac{1}{2}^+\right\rangle\nonumber\\
\left|1^+,j_l^P=\frac{3}{2}^+\right\rangle
\end{array}
\right )=\left(\begin{array}{cc} \cos\theta&\sin\theta\\
                                 -\sin\theta&\cos\theta\end{array}\right)
\left (\begin{array}{c}
 \left|^1P_1\right\rangle\nonumber\\
\left|^3P_1\right\rangle
\end{array}
\right )\,,
\end{eqnarray}
where $\theta$ is the mixing angle with
$\theta=-\tan^{-1}\sqrt{2}=-54.7^\circ$ based on the estimate in
the heavy quark limit. However, one can not determine the exact
value of $\theta$ when $m_Q$ is finite. In Ref. \cite{zhu-dai},
Dai and Zhu indicated that there does not exist a large difference
between the value of $\theta$ for the case of $m_Q\to finity$ and
that for the case of $m_Q\to \infty$.

By the $^3P_0$ model, we obtain a general relationship between
S-wave (D-wave) decay amplitude of $s\bar{b}(^1P_1)\to B^*\bar{K}$
and that of $s\bar{b}(^3P_1)\to B^*\bar{K}$
\begin{eqnarray}
&&\left (\begin{array}{c}
 \mathcal{M}\left[|s\bar{b}(^1P_1)\rangle\to |B^*\bar
 K\rangle_{S-wave}\right]\\\\
\mathcal{M}\left[|s\bar{b}(^1P_1)\rangle\to |B^*\bar
K\rangle_{D-wave}\right]
\end{array}
\right )\nonumber\\&&\quad\quad\quad=\left(\begin{array}{cc}
-\frac{1}{\sqrt{2}}&0\\\\
                                 0&\sqrt{2}\end{array}\right)
\left (\begin{array}{c}
 \mathcal{M}\left[|s\bar{b}(^3P_1)\rangle\to |B^*\bar
 K\rangle_{S-wave}\right]\\\\
\mathcal{M}\left[|s\bar{b}(^3P_1)\rangle\to |B^*\bar
K\rangle_{D-wave}\right]\end{array} \right )\,.
\end{eqnarray}
Further the amplitude squared of $1^+\to B^*\bar{K}$ decay can be
expressed as
\begin{eqnarray}
&&\left\{\begin{array}{l} |M[1^+(S)\to(B^* \bar{K})_{S-wave}]|^2\\
|M[1^+(S)\to(B^* \bar{K})_{D-wave}]|^2\\|M[1^+(T)\to(B^*
\bar{K})_{S-wave}]|^2\\|M[1^+(T)\to (B^* \bar{K})_{D-wave}]|^2
\end{array}\right.
 \propto\left\{\begin{array}{l}
(\cos\theta-{\sqrt{2}}\sin\theta)^2|A_{S}|^2\\
(\cos\theta+\frac{1}{\sqrt{2}}\sin\theta)^2|A_{D}|^2\\
(-\sin\theta-{\sqrt{2}}\cos\theta)^2|A_{S}|^2\\
(-\sin\theta+\frac{1}{\sqrt{2}}\cos\theta)^2|A_{D}|^2\end{array}\right.\label{relation}
\end{eqnarray}
with $A_{S(D)}=\mathcal{M}\left[|s\bar{b}(^1P_1)\rangle\to
|B^*\bar
 K\rangle_{S(D)-wave}\right]$.

\begin{center}
 \begin{figure}[htb]
  \begin{tabular}{cc}
  \scalebox{0.8}{\includegraphics{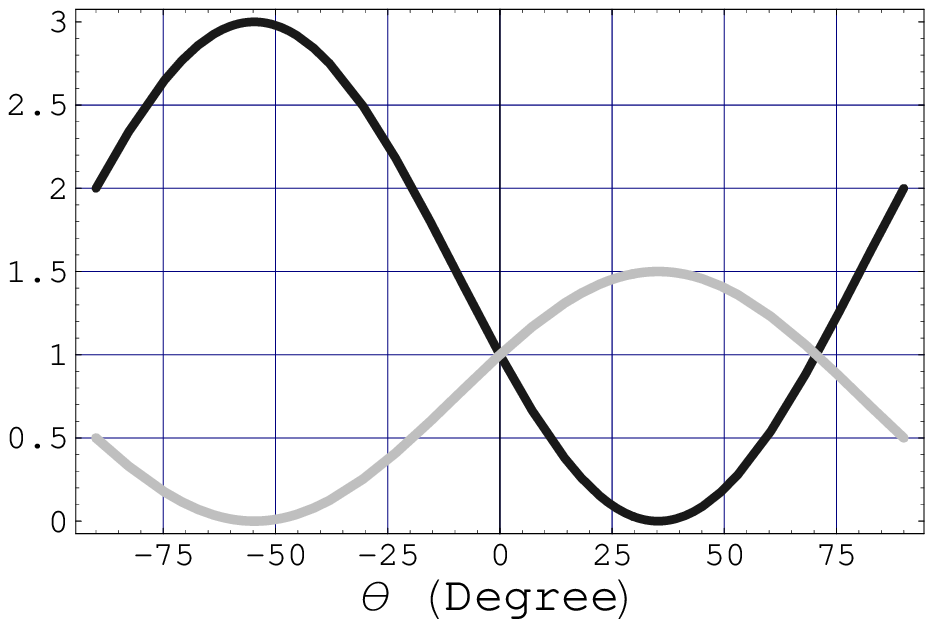}}&
  \scalebox{0.8}{\includegraphics{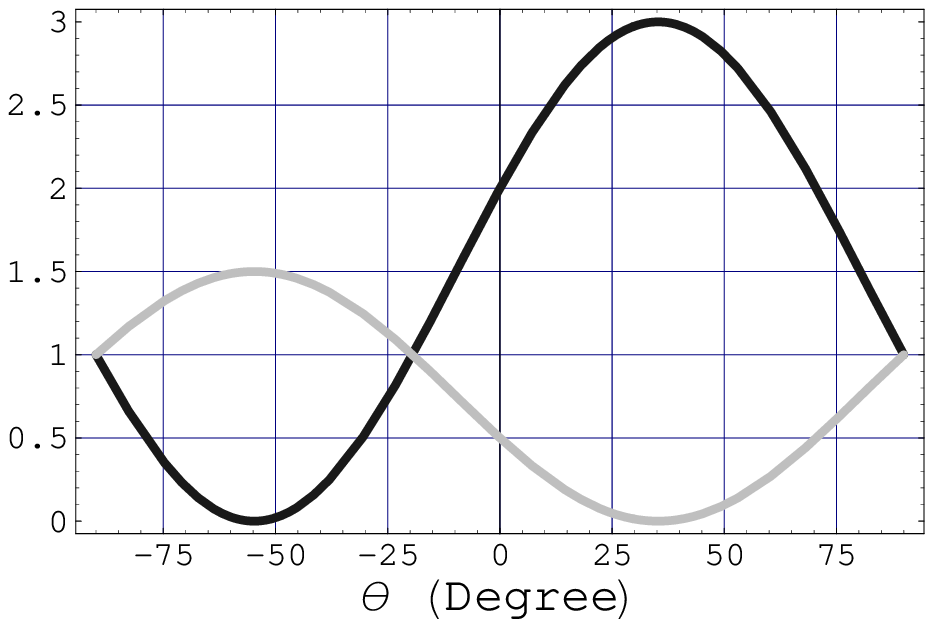}}\\(a)&(b)\\
  \end{tabular}
  \caption{The dependence of the factor in front of $|A_{S,D}|^2$ of
   Eq. (\ref{relation}) on $\theta$. The black and grey lines in
   both of the diagrams correspond to S-wave and D-wave decays,
   respectively. Here diagrams (a) and (b) are the results of
   $1^{+}$ states in $S$ and $T$ doublets, respectively.\label{S-D}}
 \end{figure}
\end{center}

In Fig. \ref{S-D}, one shows the variation of the factor in front
of $|A_{S(D)}|^2$ of Eq. (\ref{relation}) to the mixing angle
$\theta$. For the case of the decay of $1^+$ state in $S$ doublet,
$1^+$ state mainly decays into $B^*\bar{K}$ by the S-wave
amplitude since there exists the constructive (destructive)
interference between the S-wave (D-wave) decay amplitudes of
$|^1P_1\rangle$ and $|^3P_1\rangle$ states when taking
$\theta=-54.7^{\circ}$. On the contrary, for the case of $1^+$
state in $T$ doublet, the D-wave decay amplitude play the dominant
role for the decay of $1^+$ state into $B^*\bar{K}$ since the
effect of the interference between the S-wave (D-wave) decay
amplitudes of $|^1P_1\rangle$ and $|^3P_1\rangle$ states is
contrary to that of $1^+$ state in $S$ doublet when taking
$\theta=-54.7^{\circ}$. This is the reason for the total widths of
$1^+$ states existing in $S$ and $T$ doublets being wide and
narrow respectively.

\begin{center}
 \begin{ruledtabular}
  \begin{table}
   \begin{tabular}{c|c|cc}
    Mode&$(J,L)$&Decay amplitude\\\hline
    $B_{s1}(5830)^0\to$&(1,0)&$\frac{\sqrt{2}(-\sin\theta-\sqrt{2}\cos\theta)\gamma\sqrt{E_{A}E_B E_C}}{9}[I_0-2I_\pm]$\\
    $B^{*+}K^-$&(1,2)&$\frac{-{2}(-\sin\theta+1/\sqrt{2}\cos\theta)\gamma\sqrt{E_{A}E_B
    E_C}}{9}[I_0+I_\pm]$\\\hline
    $B_{s2}^*(5840)^0\to$&&\\
    $B^+K^-$&(0,2)&$\frac{-{2}}{3\sqrt{15}}\gamma\sqrt{E_{A}E_B
    E_C}[I_0+I_\pm]$\\\hline
    $B_{s2}^*(5840)^0\to$&&\\
    $B^{*+}K^-$&(1,2)&$\frac{-\sqrt{2}}{3\sqrt{5}}\gamma\sqrt{E_{A}E_B E_C}[I_0+I_\pm]$\\
   \end{tabular}\caption{The decay amplitude of the two-body strong decays of
                 $B_{s1}(5830)^0$ and $B^{*}_{s2}(5840)^0$. Here functions $I_{\pm,0}$
                 are listed in the appendix.  \label{two-body}
                }
  \end{table}
 \end{ruledtabular}
\end{center}

In Table \ref{two-body}, one presents the two-body decay
amplitudes of $B_{s1}(5830)^0$ and $B_{s2}^*(5840)^0$ calculated
by the $^3P_0$ model. The values of the parameters involved in the
$^3P_0$ model include the strength of the quark pair creation from
the vacuum and the $R$ value in the HO wave function listed in
Table \ref{parameter}. As a dimensionless parameter in the
$^{3}P_0$ model, $\gamma$ is taken as $6.9$ \cite{Godfrey}, which
is $\sqrt{96\pi}$ times larger than that used by the other groups
\cite{close-3p0,kokoski}. The $R$ value in the HO wave function
can be fixed to reproduce the realistic root mean square (RMS)
radius by solving the schr\"{o}dinger equation with the linear
potential \cite{godfrey}.

\begin{center}
 \begin{ruledtabular}
  \begin{table}
    \begin{tabular}{c||ccccc} &mass (MeV) \cite{PDG}&$R$ (GeV$^{-1}$) \cite{godfrey}\\\hline
    $B_{s1}(5830)$&5829.4&1.79\\
    $B_{s2}^*(5840)$&5839.7&1.92\\
    $B$&5279.2&1.59\\
    $B^*$&5325.1&1.75\\
    $B_{s}$&5336.3&1.45\\
    $B_{s}^*$&5412.8&1.59\\
    $K$&493.7&1.41 \\
    $f_0(980)$&980.0&2.00\\
    \end{tabular}\caption{The parameters relevant to the two-body strong decays of
                          $B_{s1}(5830)^0$ and $B^{*}_{s2}(5840)^0$ in the $^3P_0$ model \cite{PDG,godfrey}.
                          \label{parameter}
                 }
    \end{table}
 \end{ruledtabular}
\end{center}

\begin{center}
\begin{figure}[htb]
\begin{tabular}{cc}
\scalebox{0.8}{\includegraphics{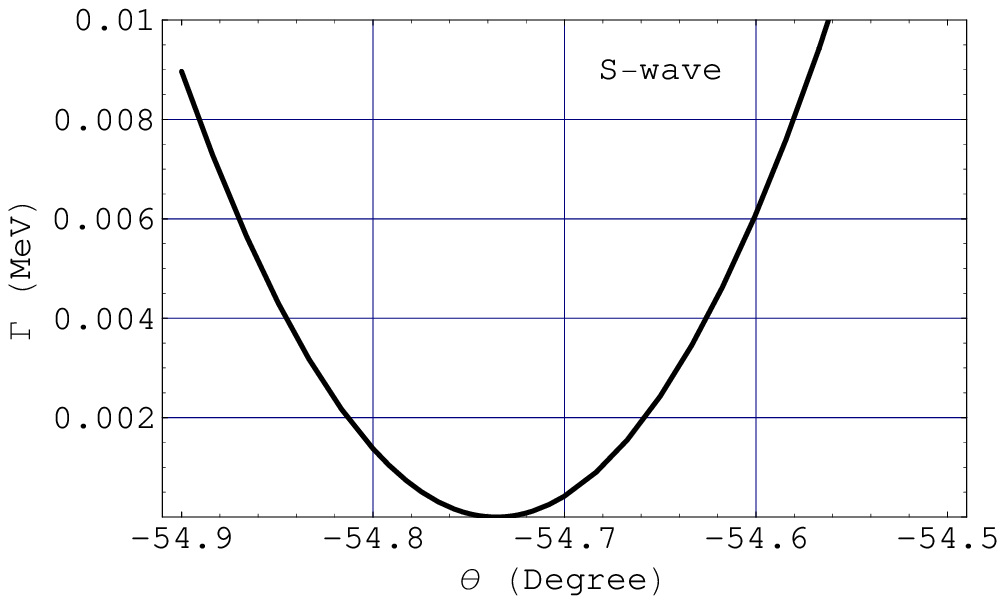}}&
\scalebox{0.8}{\includegraphics{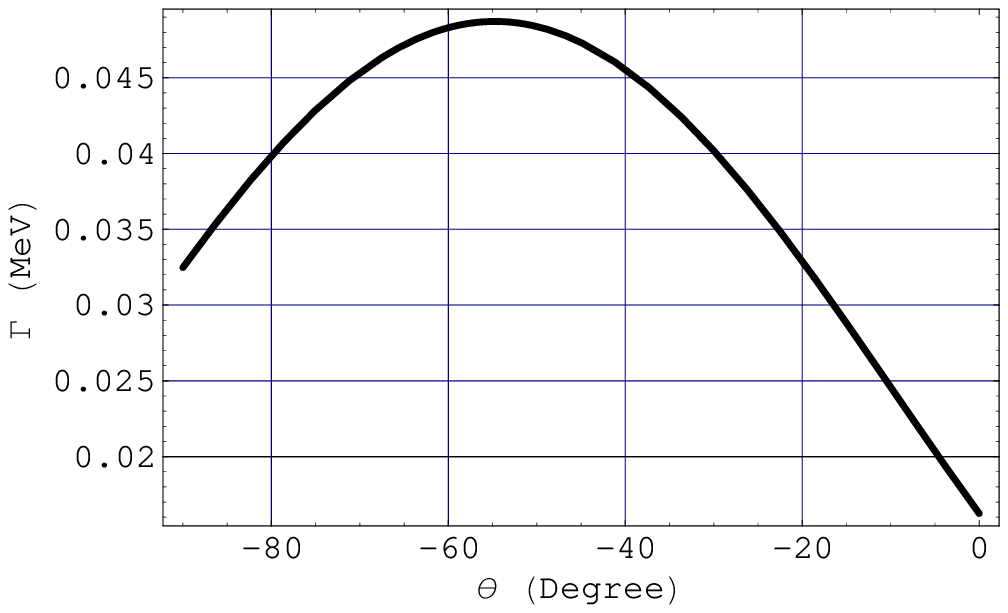}}\\(a)&(b)\\
\end{tabular}
\caption{The dependence of the partial decay width of
$B_{s1}(5830)^0\to B^{*+}K^-$ on the mixing angle $\theta$. Here
(a) and (b) respectively corresponds to S-wave and D-wave decay
widths. \label{FF-1}}
\end{figure}
\end{center}
\begin{center}
\begin{figure}[htb]
\begin{tabular}{cc}
\scalebox{0.8}{\includegraphics{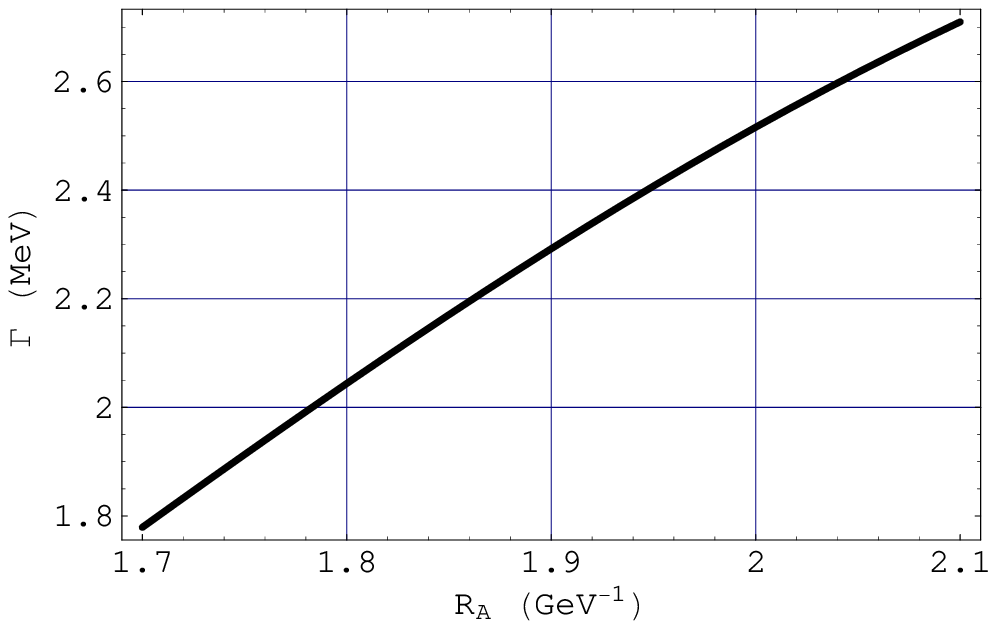}}&
\scalebox{0.8}{\includegraphics{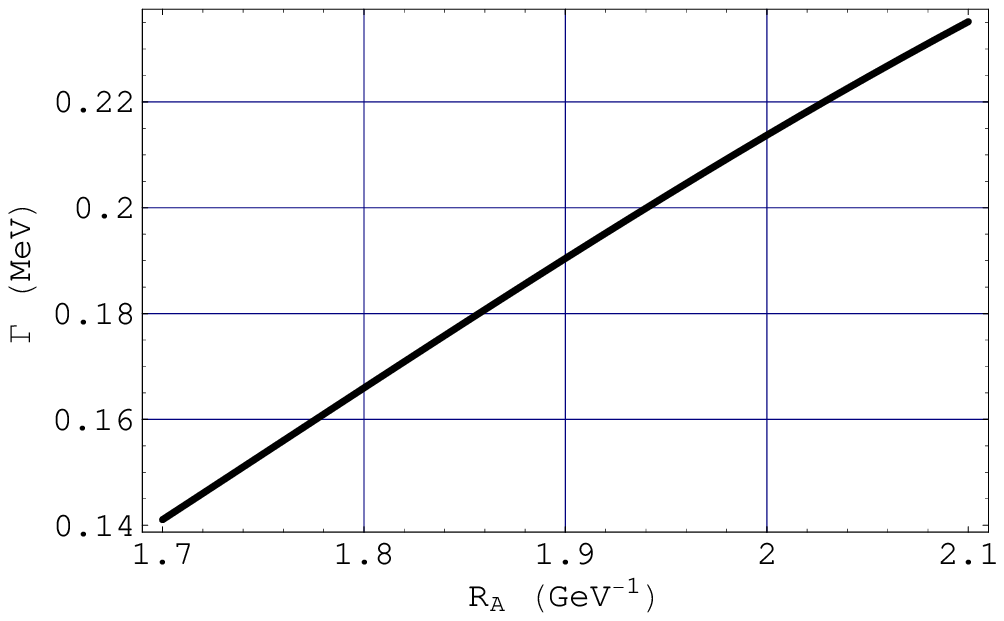}}\\(a)&
(b)\\
\end{tabular}
\caption{The variation of the two-body decay for (a)
$B_{s2}^*(5840)^0\to B^{+}K^-$ and (b) $B_{s2}^{*}(5840)^0\to
B^{*0}K^- $ with the factor $R$ of the HO wavefunction of
$B_{s2}^*(5840)^0$. \label{FF-2}}
\end{figure}
\end{center}

In Fig. \ref{FF-1} and \ref{FF-2}, one shows the dependence of the
decay width of $B_{s1}(5830)^0\to B^{*+}K^-$ on the mixing angle
$\theta$ and the variations of the decay widths of
$B_{s2}^{*0}(5840)\to B^{0}K^-,B^{*0}K^- $ to the length factor
$R$ of HO wavefunction of $B_{s2}^{*0}(5840)$.

As indicated in Fig. \ref{FF-2}, to some extent, the result
obtained by the $^3P_0$ model is sensitive to $R$ value of HO
wavefunction. Since $R$ values can be determined by reproducing
the realistic root mean square (RMS) radius when solving the
schr\"{o}dinger equation with the linear potential \cite{godfrey},
thus we fix the $R$ as the values listed in Table
\ref{parameter}, and obtain 
the partial wave decay width and the two-body decay width of
$B_{s1}(5830)^0\to B^{*+}K^-$ and $B_{s2}^{*0}(5840)\to
B^{0}K^-,B^{*0}K^- $, which are listed in Table
\ref{width-two-body}. The numerical result of $B_{s1}(5830)^0\to
B^{*+}K^-$ indicates that the S-wave partial wave decay width can
be ignored comparing with that of the D-wave when taking
$\theta=-54.7^\circ$, which is consistent with the result in quark
model.
\begin{center}
\begin{ruledtabular}
\begin{table}
\begin{tabular}{c|l|cc}
Mode&$\Gamma_{JL}$ (MeV)&$\Gamma_{two-body}$ (MeV)\\\hline
$B_{s1}(5830)^0\to$&$\Gamma_{10}=4.2\times10^{-4}$&\\
$B^{*+}K^-$&$\Gamma_{12}=4.9\times10^{-2}$&$4.9\times10^{-2}$\\\hline
$B_{s2}^*(5840)^0\to$&&\\
$B^+K^-$&$\Gamma_{02}=2.3$&2.3\\\hline
$B_{s2}^*(5840)^0\to$&&\\
$B^{*+}K^-$&$\Gamma_{12}=0.2$&0.2\\
\end{tabular}
\caption{The decay widths of two-body strong decays of
$B_{s1}(5830)^0$ and $B^{*}_{s2}(5840)^0$. Here one takes
$\theta=-54.7^\circ$ for $B_{s1}(5830)^0$ decay and adopts the $R$
values listed in Table \ref{parameter}.  \label{width-two-body}}
\end{table}
\end{ruledtabular}
\end{center}

In Table \ref{compare}, we further compare our numerical results
of the two-body strong decays of $B_{s1}(5840)$ and
$B_{s2}^*(5840)$ with the theoretical values calculated by the
other models. For the $B_{s1}(5830)^0\to B^{*}\bar{K}$ decay rate,
our result is far smaller than that from Ref. \cite{colangelo} and
is the same order of magnitude as those of Refs. \cite{FM,zhong}.
The rates of $B_{s2}(5840)\to B\bar{K}$ process predicted by the
different models are consistent with each other at the order of
magnitude. For the result of $B_{s2}(5840)\to B^*\bar{K}$, one
finds that there exists a big difference between the rate from
Ref. \cite{colangelo} and that from our calculation while the
results in Refs. \cite{EHQ,zhong} are consistent with our result.
Thus, we expect the experimental measurement of the two-body decay
rates of $B_{s1}(5840)$ and $B_{s2}^*(5840)$, which will be
helpful not only for clarifying the mist but also for further
testing the different effective models. One also notices that the
ratio of $\Gamma(B_{s2}^*(5840)\to B^*\bar{K})$ to
$\Gamma(B_{s2}^*(5840)\to B\bar{K})$ can provide the useful
information to test the model. In this work, we obtain
$$\zeta=\frac{\Gamma(B_{s2}^*(5840)\to
B^*\bar{K})}{\Gamma(B_{s2}(5840)\to B\bar{K})}\sim 8.7\%,$$ which
is close to the value $6\%$ from the chiral quark model in Ref.
\cite{zhong}. The results shown in diagrams (a) and (b) of Fig.
\ref{FF-2} also indicate the ratio $\zeta$ is a constant
basically, which is not varied with the $R$ value in the HO wave
function to some extent.

\begin{center}
\begin{ruledtabular}
\begin{table}
\begin{tabular}{c|cccccccc}
Mode&$\Gamma$ \cite{godfrey}&$\Gamma$ \cite{EHQ}&$\Gamma$
\cite{FM}&$\Gamma$ \cite{colangelo}&$\Gamma$ \cite{zhong}&this
work\\\hline
$B_{s1}(5830)\to B^*\bar{K}$&-&$<1$&0.28&$3.5$&$0.4\sim 1$&0.098\\
$B_{s2}^*(5840)\to B\bar{K}$&2.6(1.9)&$1$&$7\pm 3^{\S}$&8&$2$&4.6\\
$B_{s2}^*(5840)\to B^*\bar{K}$&0.07(0.05)&$<1$&&3.2&$0.12$&0.4\\
\end{tabular}
\caption{The comparison between our results of the two-body strong
decays of $B_{s1}(5840)$ and $B_{s2}^*(5840)$ and the results
obtained by the other theoretical groups. Here all of the results
are in units of MeV. For the values with and without bracket
listed in the second column are from the calculation results of
the pseudoscalar emission model and the flux-tube-breaking model,
respectively \cite{godfrey}. $^{\S}$ Here $7\pm 3$ MeV is the
width sum over the two processes $B_{s2}^*(5840)\to
B\bar{K},\,B^*\bar{K}$ \cite{FM}. \label{compare}}
\end{table}
\end{ruledtabular}
\end{center}

\section{Double pion decays}\label{sec4}

\begin{center}
\begin{figure}[htb]
\scalebox{0.8}{\includegraphics{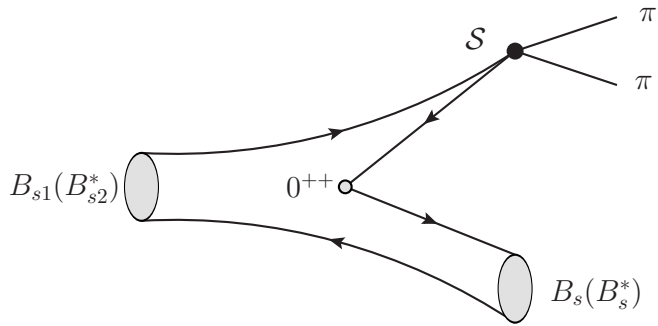}} \caption{The
double pion decay of $B_{s1}(5830)$ and $B_{s2}^*(5840)$ via the
virtual intermediate state $\sigma$ and $f_0(980)$. Here the
vertex of $B_{s1}(5830)(B_{s2}^*(5840))\to B_{s}^{(*)}\mathcal{S}$
can be depicted by the $^3P_0$ mechanism shown in Fig.
\ref{decay}. } \label{2pi-decay}
\end{figure}
\end{center}

By our calculation of the two-body strong decays of $B_{s1}(5830)$
and $B_{s2}^*(5840)$, we learn that both $B_{s1}(5830)$ and
$B_{s2}^*(5840)$ are the two states with the narrow widths. Thus,
the double pion decay of $B_{s1}(5830)$ and $B_{s2}^*(5840)$ is an
interesting topic. For estimating their double pion decays, we
assume that $B_{s1}(5830)\to B_s^{(*)}\pi\pi$ and
$B_{s2}^*(5840)\to B_s^*\pi\pi$ can occur via the intermediate
scalar state $\sigma$ and $f_0(980)$
\cite{Ishida:2001pt,BEH,lujie,xiangliu-2860}, which are depicted
by Fig. \ref{2pi-decay}. In the following, we consider the
$f_{0}(980)$ and $\sigma$ contributions to estimate the double
pion decay rates of $B_{s1}(5830)$ and $B_{s2}^*(5840)$.

The general expression of the decay width of the two pion decay of
$B_{s1}(5830)^0$ and $B_{s2}^{*}(5840)^0$ is
\begin{eqnarray}
\Gamma_{A\to B+\mathcal{S}\to
B+\pi+\pi}=\sum_{\mathcal{S}=\sigma,\,f_{0}}\frac{1}{\pi}\int_{4m_{\pi}^2}^{(M_{A}-M_B)^2}\mathrm{d}\,{r}\,
\sqrt{r}\,\frac{\Gamma_{A\to
B+\mathcal{S}}(r)\,\cdot\,\Gamma_{\mathcal{S}\to\pi+\pi}(r)
}{(r-m_{\mathcal{S}}^2)^2+(m_{\mathcal{S}}\Gamma_{S})^2},\label{general}
\end{eqnarray}
where $A$ and $B$ denote the initial and final bottom-strange
mesons in the two pion process of $B_{s1}(5830)^0$ and
$B_{s2}^{*}(5840)^0$.

The interaction of scalar state ($\mathcal{S}=f_0(980),\,\sigma$)
with the two pions is described by the effective Lagrangian
\begin{eqnarray}
\mathcal{L}_{\mathcal{S}\pi\pi}=g_{\mathcal{S}}\,
{\mathcal{S}}(2\pi^+\pi^-+\pi^0\pi^0).
\end{eqnarray}
By the total widths of $f_0(980)$ ($\Gamma_{f_0}=40\sim100$ MeV)
and $\sigma$ ($\Gamma_{\sigma}=600\sim1000$ MeV), one obtains the
values of the coupling constant $g_{f_0}=0.83\sim 1.3$ GeV and
$g_{\sigma}=2.6\sim 3.4$ GeV. Here we take $m_{\sigma}=600$ MeV
\cite{PDG}. Thus the amplitude $\Gamma_{\mathcal{S}\to \pi+\pi}$
can be expressed as
\begin{eqnarray}
\Gamma_{\mathcal{S}\to \pi+\pi}(r)=\frac{g_{\mathcal{S}}^2
{\lambda}^2}{8\pi}\frac{p_1(r)}{r}
\end{eqnarray}
with $p_{1}(r)=\sqrt{(r-4m_{\pi}^2)/4}$. ${\lambda}$ is taken as
$\sqrt{2}$ and $1$ for $\pi^+\pi^-$ and $\pi^0\pi^0$,
respectively.

One uses the $^3P_0$ model to calculate the matrix elements of the
transitions of $B_{s1}(5830)^0\to B_{s}^{(*)0}\mathcal{S}$ and
$B_{s2}^{*}(5840)^0\to B_{s}^{*0}\mathcal{S}$. Different from
$B_{s1}(5830)^0\to B^{*}\bar K$ and $B_{s2}^{*}(5840)^0\to
B\bar{K},B^{*}\bar{K}$ decays discussed in Sec. \ref{sec3},
$B_{s1}(5830)^0\to B_{s}^{(*)0}\mathcal{S}$ and
$B_{s2}^{*}(5840)^0\to B_{s}^{*0}\mathcal{S}$ are not only the
P-wave decays with $L=1$, but also are relevant to the $s\bar{s}$
quark pair creation. The strength of $s\bar{s}$ creation satisfies
$\gamma_{s}=\gamma/\sqrt{3}$ \cite{yaouanc-1} due to the flavor
dependence of the strength of quark pair creation
\cite{yaouanc-1,strangness}. The relevant transition elements are
shown in Table \ref{three-body}. The factor $\alpha_{\mathcal{S}}$
in Table \ref{three-body} is from the flavor wavefunction of
$\sigma$ and $f_0(980)$
\begin{eqnarray}
\sigma&=&\frac{1}{\sqrt{3}}(u\bar{u}+d\bar{d}+s\bar{s}),\\
f_0&=&\Big(\frac{u\bar{u}+d\bar{d}}{\sqrt{2}}\Big)\cos\varphi+s\bar{s}\sin\varphi,
\end{eqnarray}
where $\varphi=-48^\circ\pm 6^{\circ}$ due to the observation of
$f_0(980)\to \gamma\gamma$ decay mode \cite{f0}. Here
$\alpha_{\sigma}=1/\sqrt{3}$ and $\alpha_{f_0}=\sin\varphi$. Using
eq. (\ref{de}), we obtain $\Gamma_{A\to B+\mathcal{S}}(r)$ in eq.
(\ref{general}).

\begin{center}
\begin{ruledtabular}
\begin{table}
\begin{tabular}{c|lcc}
Mode&\quad\quad\quad\quad Decay amplitude\\\hline
$B_{s1}(5830)^0\to$&$\alpha_{\mathcal{S}}\big\{-\frac{\sqrt{2}}{9}\gamma_s\sqrt{E_A
E_B E_C}(2I^{0,-1}_{-1,0}+I^{0,0}_{0,0}
)(-\sin\theta)$\\
$B_s^0 \mathcal{S}$&$-\frac{2}{9}\gamma_s\sqrt{E_A E_B
E_C}(I^{-1,0}_{-1,0}+I^{-1,1}_{0,0})\cos\theta\big\}$\\\hline
$B_{s1}(5830)^0\to$&$\alpha_{\mathcal{S}}\big\{
\frac{2}{9}\gamma_s\sqrt{E_A E_B E_C}(I^{-1,0}_{-1,0}+I^{-1,1}_{0,0})(-\sin\theta)$\\
$B_s^{*0} \mathcal{S}$&$+\frac{\sqrt{2}}{9}\gamma_s\sqrt{E_A E_B
E_C}(I^{-1,0}_{-1,0}+I^{-1,1}_{0,0}+2I^{0,-1}_{-1,0}$\\
&$+I^{0,0}_{0,0})\cos\theta\big\}$\\\hline
$B_{s2}^*(5840)^0\to$&$-\alpha_{\mathcal{S}}\frac{\sqrt{2}}{9}\gamma_s\sqrt{E_A
E_B E_C}[
I^{-1,0}_{-1,0}+I^{-1,1}_{0,0}-2I^{0,-1}_{-1,0}$\\
$ B_{s}^{*0} \mathcal{S}$&$-I^{0,0}_{0,0}]$
\end{tabular}
\caption{The decay amplitude of the three-body strong decays of
$B_{s1}(5830)^0$ and $B^{*}_{s2}(5840)^0$. Here functions
$I_{M_{L_B},M_{L_C}}^{M_{L_A},m}$ are listed in the appendix.
\label{three-body}}
\end{table}
\end{ruledtabular}
\end{center}

\begin{center}
\begin{figure}[htb]
\begin{tabular}{cc}
\scalebox{0.8}{\includegraphics{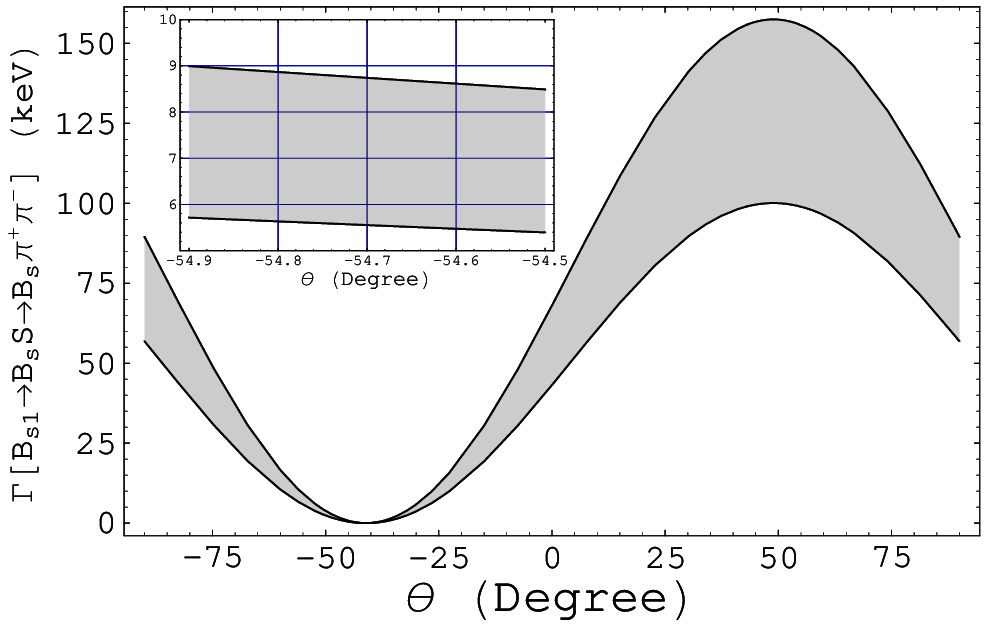}}&
\scalebox{0.8}{\includegraphics{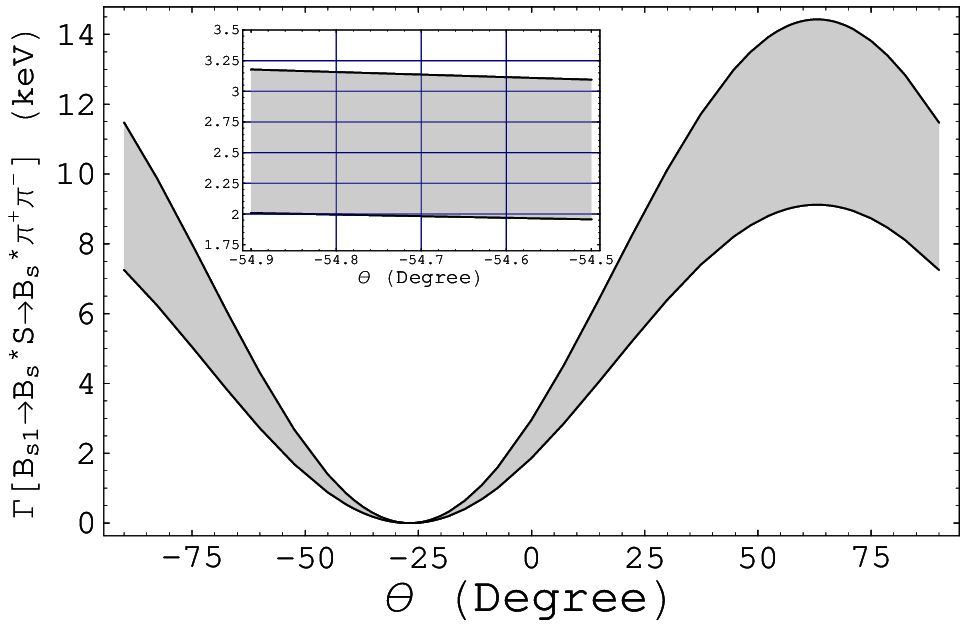}}\\(a)&(b)\\
\end{tabular}
\caption{(a) The variation of the decay width of
$B_{s1}(5830)^0\to B_{s}^0\pi^+\pi^-$ with the mixing angle
$\theta$, $g_{f_0}=0.83\sim 1.3$ GeV and $g_{\sigma}=2.6\sim 3.4$
GeV; (b) For the case of $B_{s1}(5830)^0\to B_{s}^{*0}\pi^+\pi^-$.
In the left-top diagrams of both (a) and (b), we show the enlarged
detail around $\theta=-54.7^\circ$. \label{FF-4}}
\end{figure}
\end{center}

\begin{center}
\begin{figure}[htb]
\begin{tabular}{c}
\scalebox{1}{\includegraphics{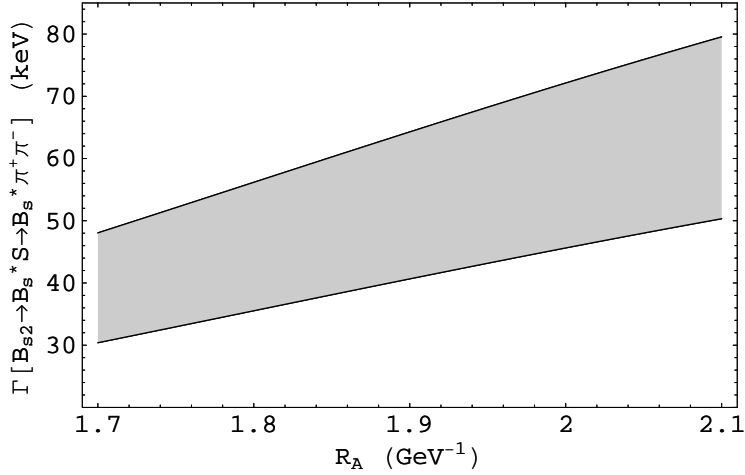}}
\end{tabular}
\caption{The dependence of decay width of $B_{s2}^{*}(5840)^0\to
B_s^{*0}\pi^+\pi^-$ on the $R$ value of the HO wavefunction of
$B_{s2}^*(5840)^{0}$, $g_{f_0}=0.83\sim 1.3$ GeV and
$g_{\sigma}=2.6\sim 3.4$ GeV. \label{FF-5}}
\end{figure}
\end{center}

In Fig. \ref{FF-4}, the dependence of the double pion decay of
$B_{s1}(5830)^0$ on the mixing angle is given. We also present the
variation of $B_{s2}^{*}(5840)^0\to B_s^{*0}\pi^+\pi^-$ with the
parameter $R$ of the HO wavefunction of $B_{s2}^{*}(5840)^0$ in
Fig. \ref{FF-5}. Here the shadow in Figs. \ref{FF-4} and
\ref{FF-5} is the possible value of the decay width. The decay
width of the double pion strong decays of $B_{s1}(5830)^0$ and
$B^{*}_{s2}(5840)^0$ are shown in Table \ref{width-two-pion} when
taking the mixing angle $\theta=-54.7^\circ$ and the $R$ value
listed in Table \ref{parameter}.

\begin{center}
\begin{ruledtabular}
\begin{table}
\begin{tabular}{c|ccc|ccc}
Mode&$\Gamma_{\pi^+\pi^-}$&$\Gamma_{\pi^+\pi^-}^{\sigma}$
&$\Gamma_{\pi^+\pi^-}^{f_0(980)}$ &$\Gamma_{\pi^0\pi^0}$
&$\Gamma_{\pi^0\pi^0}^{\sigma}$ &$\Gamma_{\pi^0\pi^0}^{f_0(980)}$
\\\hline
$B_{s1}(5830)^0\to B_s^{0}\pi\pi$&$5.6\sim 8.7$&$1.0\sim1.1$&$4.5\sim7.6$&$6.1\sim9.6$&$1.1\sim1.2$&$4.9\sim8.3$\\
$B_{s1}(5830)^0\to B_s^{*0}\pi\pi$&$2.0\sim3.1$&$0.37\sim0.41$&$1.6\sim2.7$&$2.3\sim3.7$&$0.44\sim0.49$&$1.9\sim3.2$\\
$B_{s2}^*(5840)^0\to B_s^{*0}\pi\pi$&$41.7\sim 65.9$&$7.8\sim8.7$&$33.8\sim57.2$&$48.1\sim76.1$&$9.1\sim10.1$&$39.1\sim66.1$\\
\end{tabular}
\caption{The decay widths of the double pion strong decays of
$B_{s1}(5830)^0$ and $B^{*}_{s2}(5840)^0$. Here one takes
$\theta=-54.7^\circ$ for $B_{s1}(5830)^0$ decay, and fixes all
$R$'s with the typical values listed in Table \ref{parameter}.
$\Gamma^{\sigma}_{\pi\pi}$ and $\Gamma^{f_0(980)}_{\pi\pi}$ show
the separate contributions of $\sigma$ and $f_0(980)$. All results
are in units of keV. \label{width-two-pion}}
\end{table}
\end{ruledtabular}
\end{center}

\section{Short summary}\label{sec5}

In this work, we study the two-body strong decays and the double
pion decays of the newly observed $B_{s1}(5830)$ and
$B_{s2}^*(5840)$ in the framework of the $^3P_0$ model. Our result
shows that the two-body strong decay widths of $B_{s1}(5830)$ and
$B_{s2}^*(5840)$ are about 98 keV and 5.0 MeV, respectively, when
we choose the fixed parameter presented in Sec. \ref{sec3}.
$B_{s1}(5830)$ and $B_{s2}^*(5840)$ are of narrow decay widths,
which is due to the limitation of phase space and the domination
of D-wave decay for the decays of $B_{s1}(5830)$ and
$B_{s2}^*(5840)$ into $B\bar K$ and $B^*\bar{K}$. Since the
two-body strong decay is the dominant decay mode for
$B_{s1}(5830)$ and $B_{s2}^*(5840)$, thus one expects that the
total decay widths of $B_{s1}(5830)$ and $B_{s2}^*(5840)$ are
almost not far away from their two-body decay widths at the order
of magnitude.

We also calculate the double pion decay of $B_{s1}(5830)$ and
$B_{s2}^*(5840)$ by assuming the double pion from $\sigma$ and
$f_0(980)$. The double pion decay widths are of the order of a few
keV and up to the order of magnitude of a few tens of keV for
$B_{s1}(5830)$ and $B_{s2}^*(5840)$, respectively. Although the
double pion decay widths of $B_{s1}(5830)$ and $B_{s2}^*(5840)$
are smaller than those of their two-body strong decay, the double
pion decay rates of $B_{s1}(5830)$ and $B_{s2}^*(5840)$ are
sizable. Thus we suggest future experiments to search the double
pion decay mode of $B_{s1}(5830)$ and $B_{s2}^*(5840)$.

Up to now, the experimental values of the total width of
$B_{s1}(5830)$ and $B_{s2}^*(5840)$ have not been given. To some
extent, our study is instructive for finally determining the total
width of the two newly observed $B_s$ meson in the following
experiments. Of course it is also a good way to further test the
$^3P_0$ model and other effective models.


\section*{Acknowledgments}

We thank Prof. Shi-Lin Zhu for his suggestions. L.X. would like to
thank Dr. Xian-Hui Zhong for useful communication. Z.G.L is
support by National Natural Science Foundation of China under
Grants 10625521 and 10721063 and Ministry of Education of China.
X.L. is supported by \emph{Funda\c{c}\~{a}o para a Ci\^{e}ncia e a
Tecnologia of the Minist\'{e}rio da Ci\^{e}ncia, Tecnologia e
Ensino Superior} of Portugal (SFRH/BPD/34819/2007) and National
Natural Science Foundation of China under Grants 10705001.

\section*{Appendix}

When $L_A=1$ and $L_B=L_C=0$, the spatial overlap
$I_{M_{L_B},M_{L_C}}^{M_{L_A},m}$ is simplified as
$\delta^3(\textbf{K}_B+\textbf{K}_C)I_{m'n'}(\textbf{K})$, where
\begin{eqnarray}
I_{m'n'}(\textbf{K})&=&\left(i\frac{\sqrt{2}}
{\pi^{3/4}}\right)\left(\frac{1}{\pi^{3/4}}\right)^{2}
\left(\frac{1}{2}\right)^3\,\left(\frac{3}{4\pi}\right)^{1/2}R_A^{5/2}R_B^{3/2}
R_C^{3/2}\exp\left(-\frac{1}{8}\zeta^2\textbf{K}^2\right)\nonumber\\
&&\times\int{\rm
d}\mathbf{k}\left[-k_{m'}k_{n'}+(1-\eta^2)K_{m'}K_{n'}\right]\,
\exp\left(-\frac{1}{8}\Delta^2\mathbf{k}^2\right).
\end{eqnarray}
The parameters $\Delta$, $\zeta$ and $\eta$ are defined as
\begin{eqnarray*}
&&\Delta^2=R_A^2+R_B^2+R_C^2, \,\,
\eta=\frac{R_A^2+\xi_1R_B^2+\xi_2R_C^2}{R_A^2+R_B^2+R_C^2},\nonumber\\
&&\zeta^2=R_A^2+\xi_1^2R_B^2+\xi_2^2R_C^2-\frac{(R_A^2+\xi_1R_B^2+\xi_2R_C^2)^2}{R_A^2+R_B^2+R_C^2}.\nonumber\\
\end{eqnarray*}
The $\xi_1$ and $\xi_2$ represent the mass difference effects in
mesons
\begin{eqnarray*}
\xi_1=\frac{m_3-m_1}{m_3+m_1}, \quad
\xi_2=\frac{m_4-m_2}{m_4+m_2}, \quad m_3=m_4.
\end{eqnarray*}
Here $m_i$ denotes the quark mass. In this work, we take
$m_{u}=m_d=0.22$ GeV, $m_s=0.419$ GeV, $m_b=1.977$ GeV
\cite{godfrey}.

The concrete calculations of the integration are trivial. After
choosing the direction of $\textbf{K}$ along $z$ axis, we obtain
the expressions $I_{\pm,0}$ in Table \ref{two-body}
\begin{eqnarray*}
&&I_{\pm}=I_{1-1}=I_{-11}\nonumber\\
&&\quad=i\frac{8\sqrt{3}}{\pi^{5/4}\Delta^5}
\left(R_A^{5/2}R_B^{3/2}R_C^{3/2}\right)\exp\left(-\frac{1}{8}\zeta^2\textbf{K}^2\right),\nonumber\\
&&I_0=I_{00}=i\frac{8\sqrt{3}}{\pi^{5/4}\Delta^5}\left(R_A^{5/2}R_B^{3/2}R_C^{3/2}\right)\nonumber\\
&&\quad\times\exp\left(-\frac{1}{8}\zeta^2\textbf{K}^2\right)\left[-1+\frac{1}{4}(1-\eta^2)\Delta^2\textbf{K}^2\right].
\end{eqnarray*}

When $L_A=L_B=1 $ and $L_C=0$, the spatial overlaps are of the
form $\delta^3(\textbf{K}_B+\textbf{K}_C)
\,I^{m',n'}_{\ell',0}(\textbf{K})$. Here $I^{m',n'}_{\ell',0}$ is
abbreviated as $I_{m'n'\ell'}$ with definition
\begin{eqnarray}
&&I_{m'n'\ell'}\nonumber\\\quad&&=\left(\frac{\sqrt{2}}{\pi^{3/4}}\right)^2
\left(\frac{1}{\pi^{3/4}}\right)\left(\frac{3}{4\pi}\right)^{1/2}
\left(\frac{1}{2}\right)^4R_A^{5/2}R_B^{5/2}R_C^{3/2}\nonumber\\
&&\quad\times\exp\left(-\frac{1}{8}\zeta^2\textbf{K}^2\right)
\int {\rm d}\mathbf{k}\,\big[-(\xi_1-\eta)k_{m'} k_{n'} K_{-\ell'}\nonumber\\
&&\quad(1+\eta)k_{m'} k_{-\ell'}K_{n'}-(1-\eta)k_{n'} k_{-\ell'}K_{m'}\nonumber\\
&&\quad+(1-\eta^2)(\xi_1-\eta)K_{m'}K_{n'}K_{-\ell'}\,\big]\exp\left(-\frac{1}{8}\Delta^2\mathbf{k}^2\right). \nonumber\\
\end{eqnarray}
The explicit results are
\begin{eqnarray}
&&I_{1-10}=I_{-110}=\frac{4\sqrt{6}}{\pi^{7/4}\Delta^5}\left(R_A^{5/2}R_B^{5/2}R_C^{3/2}\right)\,
|\textbf{K}|\exp\left(-\frac{1}{8}\zeta^2\textbf{K}^2\right)\left[-\eta+\xi_1\right],\\
&&I_{101}=I_{-10-1}=\frac{4\sqrt{6}}{\pi^{7/4}\Delta^5}\left(R_A^{5/2}R_B^{5/2}R_C^{3/2}\right)\,
|\textbf{K}|\exp\left(-\frac{1}{8}\zeta^2\textbf{K}^2\right)\left[-1-\eta\right],\\
&&I_{011}=I_{0-1-1}=\frac{4\sqrt{6}}{\pi^{7/4}\Delta^5}\left(R_A^{5/2}R_B^{5/2}R_C^{3/2}\right)\,
|\textbf{K}|\exp\left(-\frac{1}{8}\zeta^2\textbf{K}^2\right)\left[-\eta+1\right],\\
&&I_{000}=\frac{4\sqrt{6}}{\pi^{7/4}\Delta^5}\left(R_A^{5/2}R_B^{5/2}R_C^{3/2}\right)
|\textbf{K}|\exp\left(-\frac{1}{8}\zeta^2\textbf{K}^2\right)\nonumber\\
&&\quad\quad\quad\times\left[-\xi_1+3\eta+\frac{1}{4}(1-\eta^2)(\xi_1-\eta)\Delta^2\textbf{K}^2\right].
\end{eqnarray}

\end{document}